\documentclass[conference]{IEEEtran}
\IEEEoverridecommandlockouts
\usepackage{cite}
\usepackage{amsmath,amssymb,amsfonts}
\usepackage{algorithmic}
\usepackage{graphicx}
\usepackage{textcomp}
\usepackage{bbm}
\usepackage{xcolor}
\usepackage{multirow}
\def\BibTeX{{\rm B\kern-.05em{\sc i\kern-.025em b}\kern-.08em
    T\kern-.1667em\lower.7ex\hbox{E}\kern-.125emX}}
\begin{document}
\bstctlcite{IEEEexample:BSTcontrol}

\title{Experimental Analysis of Deep Hedging Using Artificial Market Simulations for Underlying Asset Simulators}

\author{
    \IEEEauthorblockN{Masanori Hirano}
    \IEEEauthorblockA{
        \textit{Preferred Networks, Inc.}\\
        Tokyo, Japan \\
        research@mhirano.jp}
}

\maketitle

\begin{abstract}
    Derivative hedging and pricing are important and continuously studied topics in financial markets.
    Recently, deep hedging has been proposed as a promising approach that uses deep learning to approximate the optimal hedging strategy and can handle incomplete markets.
    However, deep hedging usually requires underlying asset simulations, and it is challenging to select the best model for such simulations.
    This study proposes a new approach using artificial market simulations for underlying asset simulations in deep hedging.
    Artificial market simulations can replicate the stylized facts of financial markets, and they seem to be a promising approach for deep hedging.
    We investigate the effectiveness of the proposed approach by comparing its results with those of the traditional approach, which uses mathematical finance models such as Brownian motion and Heston models for underlying asset simulations.
    The results show that the proposed approach can achieve almost the same level of performance as the traditional approach without mathematical finance models.
    Finally, we also reveal that the proposed approach has some limitations in terms of performance under certain conditions.
\end{abstract}


\section{Introduction}
Derivative hedging and pricing are important matters in financial markets regarding asset risk management.
A derivative is a financial instrument whose value depends on that of an underlying asset.
It is frequently used to hedge the risk that cannot be directly hedged using only the underlying asset.
Moreover, it is important to price and hedge the risk of derivatives for security companies that issue derivatives.
Therefore, derivative pricing and hedging have been extensively studied.

Many methods exist for derivative pricing and hedging.
Traditionally, mathematical finance models, such as the Black--Scholes model \cite{black1973,merton1973}, are used for pricing and hedging.
Although these models are useful, they also have some limitations.
For example, they assume complete markets, which means that there are no frictions such as transaction costs and liquidity.
To solve this problem, some researchers have proposed a new model called deep hedging \cite{deep-hedging}.
Deep hedging is a model based on deep learning that can handle incomplete markets.
It uses neural networks to approximate the optimal hedging strategy and Monte Carlo simulations to evaluate the derivative price.
Because of these features, deep hedging can be used in more practical situations than traditional models.
In addition to those features, a notable advantage of deep hedging is its ability to use risk measure functions as objective functions.

Deep hedging is a promising model; however, there are some difficulties in selecting underlying asset simulations.
Non-stationarity exists in the actual financial markets, making it difficult to learn using only historical data.
Therefore, data augmentation via Monte Carlo simulations is used to learn about the markets in deep hedging.
However, many candidate models exist for Monte Carlo simulations of underlying assets, and it is difficult to select the best one.
For example, the geometric Brownian motion model is a simple model for the simulations; Buehler {\it et al.} \cite{deep-hedging} originally used the Heston model \cite{Heston1993}, which is a stochastic volatility model for simulations.
Harvath {\it et al.} \cite{Horvath2021} used a rough volatility model called rBergomi \cite{Bayer2015} for the simulations, and the jump-diffusion model \cite{Cox1976,Merton1976} could also be used for the simulations.
However, mathematical finance models are not always suitable because they do not consider the stylized facts of actual financial markets.
Therefore, selecting the best model for simulations is difficult.

Artificial market simulations are a promising approach for replicating stylized facts.
Artificial market simulations are financial market simulations that use multi-agent simulations.
Financial markets are highly complex because of the interactions between traders and markets, and there is no dominant equation for the phenomena occurring in financial markets.
Therefore, multi-agent simulations, in which agents can interact with each other, are necessary to reproduce well-known phenomena in financial markets.
For example, \cite{Lux1999} showed that in financial market simulations, interactions between agents (traders) were necessary to replicate common phenomena in financial markets.
In addition, \cite{CI2002} showed that the stylized facts of financial markets can be replicated using multi-agent simulations based on fundamental, chart, and noise factors.

This study uses artificial market simulations for underlying asset simulations in deep hedging.
A similar study \cite{Gao2023} merged mathematical finance models and Chiarella {\it et al.} artificial market model \cite{CI2002}.
However, no study has used artificial market simulations for underlying asset simulations in deep hedging without mathematical finance models.
Therefore, we only use artificial market simulations for underlying asset simulations in deep hedging while investigating the effectiveness of this approach.

The contributions of this study are as follows:
\begin{itemize}
    \item We propose a new approach to use artificial market simulations for underlying asset simulations in deep hedging.
    \item We investigate the effectiveness of the approach by comparing the results with those of the traditional approach.
    \item We show that the proposed approach can achieve almost the same performance as the traditional approach without mathematical finance models.
    \item But, we also show that the proposed approach has some limitations in terms of performance under some conditions.
\end{itemize}

\section{Related Works}
Mathematical finance approaches have been frequently used in terms of option hedging and prices.
The celebrated Black--Scholes model \cite{black1973,merton1973} is certainly the most famous model in the option pricing literature.
To date, the Black-Scholes model has been widely used in practice.
Its applications include deriving option pricing formulae (e.g., the Black-Scholes formula) and analyzing structures of option prices (e.g., Greeks).

Advances in neural networks have opened up a new avenue of research on nonparametric approaches to option pricing and hedging.
Hutchinson {\it et al.} \cite{Hutchinson1994} and Garcia {\it et al.} \cite{Garcia2000} showed that a multilayer perceptron (MLP) could replace the Black--Scholes model.
Malliaris {\it et al.} \cite{Malliaris1996} showed that neural networks could predict implied volatility by using the index option of the S\&P100.
For a comprehensive review, see Ruf and Wang \cite{Ruf2020} and the references therein.

In their seminal work \cite{deep-hedging,Buehler2019b}, Buehler {\it et al.} proposed a method called \emph{Deep Hedging}.
Deep hedging utilizes neural networks to model hedging strategies and construct optimal strategies by directly maximizing the utility functions derived from the final Profit \& Loss (PL).
Deep hedging has been recognized as a breakthrough in the derivative industry owing to two prominent features.
First, deep hedging can handle various types of friction using a fully computational approach.
Second, deep hedging bypasses the need for theoretical pricing models, which are prerequisites for other standard hedging methods, such as those based on Greeks.
Since the introduction of deep hedging, considerable efforts have been devoted to expanding its versatility.
Imaki {\it et al.} \cite{Imaki2023} proposed a deep hedging model that incorporated a no-transaction band \cite{Davis1993} into the neural network architecture, resulting in fast convergence of learning.
Murray {\it et al.} \cite{Murray2022-nz, Buehler2022-nr} proposed actor-critic-based reinforcement learning for deep hedging and achieved a performance almost equivalent to that of the original deep hedging.
Hirano {\it et al.} \cite{Hirano2023-iiaiaai-ndh} extended the deep hedging framework into a nested structure, thereby facilitating the use of options as hedging instruments. The authors also proposed efficient simulation designs and a learning algorithm to circumvent the computational issues.

Several recent studies focus on the data used for learning deep hedging.
In conventional deep hedging \cite{deep-hedging}, a simulator based on the Heston model \cite{Heston1993} was used for training.
Mikkila {\it et al.} \cite{Mikkila2023-gk} argued that deep hedging often depends on a specific choice of price process simulator, which could result in suboptimal performance. They suggested that improved performance could be achieved by utilizing empirical data for training. However, the amount of historical market data available is often limited.
Horvath {\it et al.} \cite{Horvath2021} proposed a deep hedging model based on rough volatility models, such as the rBergomi model \cite{Bayer2015}, which could model (non-Markov) jumps in price processes.

Among the numerous simulators of price processes, it is extremely challenging to determine which one is appropriate or whether a simulator should be used or otherwise.
This paper proposes a new approach to using artificial market simulations for underlying asset simulations in deep hedging.
\cite{Gao2023} have already proposed a new model merging mathematical finance models and Chiarella {\it et al.}'s artificial market model \cite{CI2002}.
However, the benefits of artificial market simulations are not fully utilized in their model, which still retains the features of mathematical finance models.

Originally, multi-agent simulations have been used to analyze and understand many social phenomena since around the 1970s \cite{schelling1969,schelling1969}, but they have also been employed in financial markets.
Lux et al.\cite{Lux1999} highlighted the necessity of a multi-agent simulation by demonstrating that the well-known phenomena observed in financial markets cannot be reproduced without the interaction between agents in the simulations.
Braun-Munzinger et al.\cite{Braun-Munzinger2016} constructed a multi-agent simulation for the bond market.

Social phenomena, especially in financial markets, are difficult to analyze because no dominant equation exists or has been elucidated.
Therefore, simulations are useful in the social science field \cite{Edmonds2005}.
In particular, various studies argue that multi-agent simulations are important\cite{Farmer2009,Battiston2016}.
Furthermore, Mizuta\cite{Mizuta2019} argued that multi-agent simulations in finance can contribute to financial regulation and system design.
The limitations of existing financial approaches are being argued by financial dignitaries, accelerating research on artificial market simulation.
The 2007--2008 financial crisis caused the collapse of investment banks and a global financial market network due to the default on housing loans in the U.S.
Trichet, the European Central Bank (ECB) president at the time, not only stated that traditional financial theory did not help make policy decisions during the financial crisis but also emphasized the need for behavioral economics and multi-agent simulation\cite{Trichet2011}.
Bookstaber, who has worked in risk management at investment banks and hedge funds and has also worked at the U.S. Treasury, argued for a paradigm shift to methods that could incorporate complexity, such as multi-agent simulation, in his book\cite{Bookstaber2017}, reflecting on the financial crisis.
It claims that traditional economic theory is difficult to apply in times of crisis because distortions are amplified.

Various studies have adopted artificial market simulation.
Cui et al.\cite{Cui2012} showed that it is impossible to reproduce certain phenomena observed in the stock market using only zero-intelligence agents.
Torii et al.\cite{Torii2015} conducted a simulation where price shocks spread to other stocks and analyzed the mechanism.
Mizuta et al.\cite{Mizuta2016} analyzed the impact of price quotation (minimum price unit) in the stock market using artificial markets and argued that lowering price quotation was necessary to maintain market share, thereby contributing to discussions on lowering price quotation on the Tokyo Stock Exchange.
Hirano et al.\cite{Hirano2020c} used artificial markets to analyze the impact of capital adequacy ratio regulation, showing that capital adequacy ratio regulation, which was supposed to have been introduced to ensure market stability, could amplify price shocks and suppress price increases.
Some studies reproduce flash crashes in artificial markets\cite{Leal2019,Paddrik2012}.

According to such studies, artificial market simulations are useful for analyzing and understanding financial markets.
They could also be useful for deep hedging by replicating the underlying asset price movements in actual financial markets.

\section{Task Setting}
This study considers the problem of hedging an option with its underlying asset.
In this task, an option is sold at the beginning, and hedging it using its underlying asset is required until maturity.
The option payoff construct remains unknown, and its payment risk should be reduced.

We consider a discrete-time financial market with a finite time horizon $T$ and trading dates $0=t_0 < t_1 < \cdots < t_n = T$.
At $t=t_0$, the hedger (the issuer of the option) sells one unit of the option.
The hedger aims to hedge the risk arising from the future payoff of the option by trading the underlying asset $S$.
$S$ is always assumed to be tradable, and its price at time $t$ is denoted as $S_t$.
On each trading date $t_i$, the hedger sets a new position $\delta_{t_i}$ in $S$.
This means that the trader can sell or buy an underlying asset $S$ at $t=t_0, t_1, \cdots, t_{n-1}$ and the clearance for both the option and the  underlying asset is carried at maturity $t=t_n (=T)$.
For simplification, we ignore the risk-free rate, which can be considered all values calculated as discounted if we assume no risk-free rate change occurs.
The proportional cost, whose cost coefficient for transaction prices is $c$, is charged for all transactions.
For all transactions of $S$, a proportional transaction cost is charged at $c$ per unit price.

Option $Z$ is an option whose underlying asset is $S$.
$Z$ have payoff $Z(S)$ at the maturity $t=T$.
Examples include an European option $Z(S) = \max(S_T-K, 0)$ and a Lookback option $Z(S) = \max(\max(S)- K, 0)$, where $K$ denotes the strike value of the options.
These options will also be used in our experiments.

In this study, we consider the sell-side trader of options.
This means that the trader shorts (sells) one unit of option, hedges it until maturity, and then pays the payoff $Z(S)$ to the buyer at maturity.
The Profit \& Loss (PL) of the hedger (the issuer) is formulated as follows:
\begin{eqnarray}
    \mathrm{PL}_T (Z, S, \delta) &:=& -Z(S) + (\delta\cdot S)_T - C_T (\delta)\label{eq:pl}\\
    (\delta\cdot S)_T &:=& \sum_{i=0}^{n-1} \delta_{t_i} (S_{t_{i+1}} - S_{t_i}) \\
    C_T (\delta) &:=& \sum_{i=0}^{n} cS_{t_i}|\delta_{t_i} - \delta_{t_{i-1}}|
\end{eqnarray}
where $\delta_{-1} = \delta_{t_n} = 0$.
Here, $(\delta\cdot S)_T$ is the total return resulting from the trading of the underlying asset, and $C_T (\delta)$ is the total transaction cost.

The objective function of the hedger is defined by the PL and the utility function $u$.
Specifically, the loss function to be minimized is defined as follows:
\begin{eqnarray}
    l(\delta) = - u \left(\mathrm{PL}_T(Z, S, \delta)\right).
\end{eqnarray}
For an example of the utility function $u$, entropic risk measure (ERM) is defined as
\begin{eqnarray}
    u(x) = -\frac{1}{\lambda}\log\mathbb{E}\left[\exp{(-\lambda x)}\right]\label{eq:erm}
\end{eqnarray}
where $\lambda$ is the risk preference coefficient.
Expected shortfall (conditional value at risk; CVaR) is defined as
\begin{eqnarray}
    u(x) = \frac{1}{1-\alpha} \int_{-\infty}^{\mathrm{VaR}(\alpha)} x\cdot p(x)dx\label{eq:cvar}
\end{eqnarray}
where $\alpha$ is the confidence level and satisfies $0\leq\alpha\leq 1$.

The option price $p$ is estimated using
\begin{eqnarray}
    \mathbbm{E}\left[u \left(\mathrm{PL}_T(Z, S, \delta) + p\right)\right] = u(0).
\end{eqnarray}
Here, the lower price $p$ implies a highly efficient hedging strategy.

\section{Artificial Market Simulations}
This study employed a typical artificial market simulation model based on \cite{Torii2015}.
In our simulation, 100 stylized trader agents exist, and 1 continuous double auction market exists.
On each step of the simulation, $N_{\mathrm{agent/step}}$ agents among the 100 agents are randomly selected, and they trade on the market according to the algorithm described below.
Furthermore, if the market board is thin at the beginning of the simulation, the market's liquidity will reduce, and the situation will not be similar to the actual market.
Therefore, we employ a pre-opening session before the main session to increase market liquidity.
In the following, the total number of steps over the pre-opening and main sessions is denoted as $t$

At time $t$, the stylized trader agent $i$ decides on its trading actions using the following criteria: fundamentals, chartists (trends), and noise.
Initially, the agents calculate these three factors.
\begin{itemize}
    \item Fundamental factor:
          \begin{equation}
              F_t^i = \frac{1}{\tau^{*i}} \ln{\left\{\frac{p_t^*}{p_t}\right\}}.
          \end{equation}
          where $\tau^{*i}$ is agent $i$'s mean reversion-time constant, $p_t^*$ is the fundamental price at time $t$, and $p_t$ is the price at time $t$. The fundamental price starts at 1.0 and is generated by a Brownian motion with a volatility of $\sigma^{*}$.
    \item Chartist factor:
          \begin{equation}
              C_t^i = \frac{1}{\tau^i}\sum_{j=1}^{\tau^i} r_{(t-j)} = \frac{1}{\tau^i}\sum_{j=1}^{\tau^i}\ln{\frac{p_{(t-j)}}{p_{(t-j-1)}}},
          \end{equation}
          where $\tau^i$ is the time window size of agent $i$ and $r_t$ is the logarithm of return at time $t$.
    \item Noise factor:
          \begin{equation}
              N_t^i \sim \mathcal{N} (0, \sigma).
          \end{equation}
          denotes that $N_t^i$ obeys a normal distribution with zero mean and variance $(\sigma)^2$.
\end{itemize}

Subsequently, the agents calculate the weighted averages of these three factors.
\begin{equation}
    \widehat{r_t^i} = \frac{1}{w_F^i + w_C^i + w_N^i} \left(w_F^i F_t^i + w_C^i C_t^i + w_N^i N_t^i\right),
\end{equation}
where $w_F^i, w_C^i$, and $w_N^i$ are the weights of agent $i$ for each factor.

Next, the expected price of agent $i$ is calculated using the following equation:
\begin{equation}
    \widehat{p_t^i} = p_t \exp{\left(\widehat{r_t^i} \tau^i\right)}.
\end{equation}

Subsequently, using a fixed margin of $k^i \in [k_{\mathrm{min}}, k_{\mathrm{max}}]$, we determine the actual order prices using the following rules:
\begin{itemize}
    \item If $\widehat{p_t^i} > p_t$, agent $i$ places a bid (buy order) at the price
          \begin{equation}
              \min{\left\{\widehat{p_t^i} (1-k^i), p_{t}^{\mathrm{bid}}\right\}}.
          \end{equation}
    \item If $\widehat{p_t^i} < p_t$, agent $i$ places an ask (sell order) at the price
          \begin{equation}
              \max{\left\{\widehat{p_t^i} (1+k^i), p_{t}^{\mathrm{ask}}\right\}}.
          \end{equation}
\end{itemize}
Here, $p_{t}^{\mathrm{bid}}$ and $p_{t}^{\mathrm{ask}}$ denote the best bid and ask prices, respectively.

The parameters of the artificial market simulation are tuned in the experiments.

\section{Experiments}
This section describes the experiments conducted to evaluate the effectiveness of the proposed approach using artificial market simulations for underlying asset $S$ simulations in deep hedging.
We compare the proposed approach's results with those of the traditional approach, which uses mathematical finance models such as Brownian motion and Heston models for underlying asset simulations.
Brownian motion is defined as
\begin{eqnarray}
    \Delta S_{t_i} = S_{t_i} ( \mu + \sigma \delta W_{t_i})
\end{eqnarray}
where $\delta W_{t_i}$ is an independent normal random variable and $\sigma$ is the volatility of the underlying asset.
Heston model is defined as
\begin{eqnarray}
    d S_t &=& S_t \sqrt{V_t} dW^{(1)}_t\\
    dV_t &=& \kappa (\theta - V_t) dt + \sigma \sqrt{V_t} dW^{(2)}_t
\end{eqnarray}
where $dW^{(1)}_t$ and $dW^{(2)}_t$ are the Winner processes whose correlation is $\rho$.
QE-M Method \cite{andersen2007} is used to simulate the Heston model.

As the derivatives $Z$ for this experiment, we use European and Lookback options whose maturity and strike price are 20 days and 1.0, respectively.
For the utility function $u$ ERM($\lambda = 1, 10$) and CVaR($\alpha = 0.90, 0.95, 0.99$) are used.
Experiments are carried out for those combinations of derivatives and utility functions.
For evaluation, we employed the following actual market data:
\begin{itemize}
    \item S\&P 500: American S\&P 500 index from January 2000 to August 2022
    \item S\&P 500 (old): American S\&P 500 index from January 1931 to December 1950
    \item BVSP: Brazilian Bovespa(BVSP) index from January 2000 to August 2022
\end{itemize}
Among the three datasets, S\&P 500 is used for the model development data, and the others are used for the test data.

The flow of the experiments is as follows:
\begin{enumerate}
    \item The parameters of the artificial market simulation are tunned using Optuna \cite{optuna_2019}, one of the Bayesian estimation-based hyperparameter optimization libraries. The tuning process is repeated 500 times. In each trial, the following process, that is, a) and b), is repeated:
          \begin{enumerate}
              \item generates 10,000 paths of price time series for the underlying assets using artificial market simulation, Brownian motion, or Heston model. Then, the training of deep hedging is carried out for 100 epochs.
              \item evaluate the performance of the trained deep hedging models. To get the best performances, we employed the best-performing models among the 100 epochs in terms of the pricing over S\&P 500 (old) data. Then, the best pricing is used as the objective value of tuning.
          \end{enumerate}
    \item The final evaluation is carried out using S\&P 500 and BVSP data.
\end{enumerate}

For the implementation, we used the Python--based artificial market simulation PAMS\cite{Hirano2023-pams} and the Python and pytorch--based pfhedge\cite{pfhedge} for Deep Hedging.
The deep hedging model employs moneyness, time-to-maturity, volatility, Black--Scholes Delta, and historical maximum value of moneyness (only for Lookback) as inputs, and a 32-dimensional 4-layer MLP with ReLU and Layer Normaliuzation\cite{Ba2016} was employed.
Adam was employed as an optimizer.

The parameters to be tuned are as follows and are discretized to improve interpretability.
\begin{itemize}
    \item Training parameters:\begin{itemize}
              \item Learning rate: $10^{-1}$, $10^{-2}$, $10^{-3}$, $10^{-4}$, $10^{-5}$
          \end{itemize}
    \item Brownian Motion:\begin{itemize}
              \item Mean return ($\mu$): $-0.25$, $-0.20$, ... , $0.20$, $0.25$
              \item Volatility ($\sigma$): $0.05$, $0.10$, ... , $0.45$, $0.50$
          \end{itemize}
    \item Heston Model:\begin{itemize}
              \item $\kappa$: $0$, $0.05$, ... , $0.45$, $0.50$
              \item Initial Volatility ($\sigma = \sqrt{\theta} = \sqrt{V_0}$): $0.05$, $0.10$, ... , $0.45$, $0.50$
              \item $\rho$: $-1.0$, $-0.95$, ... , $0.95$, $1.0$
          \end{itemize}
    \item Artificial Market Simulations' parameters: \begin{itemize}
              \item $N_{\mathrm{agent/step}}$: $1$, $5$, $10$
              \item $w_F, w_C$: $0.0, 1.0, 3.0, 5.0, 10.0, 30.0, 50.0$
              \item $\sigma^{*}$: $10^{-2}$, $10^{-3}$, $10^{-4}$
              \item $\sigma$: $10^{-2}$, $10^{-3}$, $10^{-4}$, $10^{-5}$
              \item $\tau^{*}_{\mathrm{min}}, \tau^{*}_{\mathrm{max}}$: $1$, $10$, ..., $10^{5}$
              \item $\tau_{\mathrm{min}}, \tau_{\mathrm{max}}$: $1$, $10$, ..., $10^{3}$
              \item $k_{\mathrm{min}}, k_{\mathrm{max}}$: $0.0$, $0.05$, ..., $0.20$
          \end{itemize}
\end{itemize}
Note that the parameters with a large or small relationship are considered the constraints of tuning.
Those parameter candidates are set based on previous studies.
For the Brownian motion and Heston models, we set the default parameters based on a previous study \cite{deep-hedging} and calculated the performance without tuning.
The default parameters used in the previous study were $\mu=0.0, \sigma=0.2, \kappa=1.0, \rho = -0.7$.

\begin{table*}[htb]
    \caption{Option pricing results in each setting}
    %\vspace{-2mm}
    \label{tab:results}
    \centering
    \begin{tabular}{ccc|cc|ccc}
        \hline
        \multicolumn{3}{c|}{Settings} & \multicolumn{2}{c|}{Default} & \multicolumn{3}{c}{After Tuning}                                                                                                                        \\
        Derivative                    & Underlying Asset Data        & Utility Function                 & Brownian            & Heston              & Artificial Market Simulation & Brownian            & Heston              \\\hline
        european                      & S\&P 500                     & ERM ($\lambda=1$)                & 0.019750            & \color{red}0.019192 & 0.019419                     & 0.019551            & 0.019299            \\
        european                      & S\&P 500                     & ERM ($\lambda=10$)               & 0.020935            & \color{red}0.020246 & 0.020845                     & 0.036039            & 0.020426            \\
        european                      & S\&P 500                     & CVaR ($\alpha=0.90$)             & 0.050233            & 0.052363            & \color{red}0.049532          & 0.061715            & 0.050061            \\
        european                      & S\&P 500                     & CVaR ($\alpha=0.95$)             & 0.061153            & 0.061949            & \color{red}0.058047          & 0.063598            & 0.060551            \\
        european                      & S\&P 500                     & CVaR ($\alpha=0.99$)             & 0.093186            & 0.095895            & 0.113233                     & \color{red}0.092216 & 0.092759            \\\hline
        european                      & BVSP                         & ERM ($\lambda=1$)                & 0.027719            & 0.027331            & \color{red}0.026537          & 0.027477            & 0.027362            \\
        european                      & BVSP                         & ERM ($\lambda=10$)               & 0.031842            & \color{red}0.031358 & 0.032121                     & 0.067403            & 0.031519            \\
        european                      & BVSP                         & CVaR ($\alpha=0.90$)             & 0.067626            & 0.070673            & \color{red}0.064339          & 0.085204            & 0.068147            \\
        european                      & BVSP                         & CVaR ($\alpha=0.95$)             & 0.079984            & 0.081648            & \color{red}0.076255          & 0.084457            & 0.080254            \\
        european                      & BVSP                         & CVaR ($\alpha=0.99$)             & \color{red}0.123999 & 0.131008            & 0.147125                     & 0.124240            & 0.126957            \\\hline
        lookback                      & S\&P 500                     & ERM ($\lambda=1$)                & 0.027260            & 0.028259            & \color{red}0.027046          & 0.027102            & 0.027245            \\
        lookback                      & S\&P 500                     & ERM ($\lambda=10$)               & 0.032112            & \color{red}0.030554 & 0.038140                     & 0.035739            & 0.030621            \\
        lookback                      & S\&P 500                     & CVaR ($\alpha=0.90$)             & 0.085309            & 0.086851            & 0.091485                     & \color{red}0.076732 & 0.082135            \\
        lookback                      & S\&P 500                     & CVaR ($\alpha=0.95$)             & 0.105119            & 0.107009            & 0.111806                     & 0.104828            & \color{red}0.095170 \\
        lookback                      & S\&P 500                     & CVaR ($\alpha=0.99$)             & 0.163010            & 0.168566            & 0.164313                     & \color{red}0.161206 & 0.161686            \\\hline
        lookback                      & BVSP                         & ERM ($\lambda=1$)                & 0.045672            & 0.047888            & \color{red}0.045210          & 0.045743            & 0.046029            \\
        lookback                      & BVSP                         & ERM ($\lambda=10$)               & 0.051926            & 0.051787            & 0.055111                     & 0.053248            & \color{red}0.051522 \\
        lookback                      & BVSP                         & CVaR ($\alpha=0.90$)             & 0.109817            & 0.106158            & 0.111802                     & \color{red}0.097037 & 0.102013            \\
        lookback                      & BVSP                         & CVaR ($\alpha=0.95$)             & 0.130552            & 0.125846            & 0.131436                     & 0.122045            & \color{red}0.115475 \\
        lookback                      & BVSP                         & CVaR ($\alpha=0.99$)             & 0.200004            & 0.20705             & 0.199006                     & 0.198911            & \color{red}0.198437 \\\hline
    \end{tabular}
\end{table*}

\section{Results and Discussion}
Table \ref{tab:results} presents the results of the options pricing experiments for each setting.
In this figure, the lower the value, the better the performance.
The red values indicate the best performance in each setting.
The results show that the proposed approach, using artificial market simulations for underlying asset simulations, can achieve almost the same performance as the traditional approach using mathematical finance models.
In particular, the proposed approach can perform best in certain settings, such as the European option with CVaR ($\alpha=0.90$) and CVaR ($\alpha=0.95$).
In addition, for options with ERM ($\lambda=1$), the proposed approach can achieve the best performance in most settings.

Additionally, a comparison of the default parameter results with the results after tuning can sometimes reveal that the default parameters achieve the best performance.
It seems that the parameter tuning was carried out using the pricing over S\&P 500 (old) data, and the parameters are sometimes only suitable for other data.
However, the default parameters are set based on the previous study and could be suitable for the recent data.

\begin{figure*}[htb]
    \centering
    \includegraphics[width=0.32\linewidth]{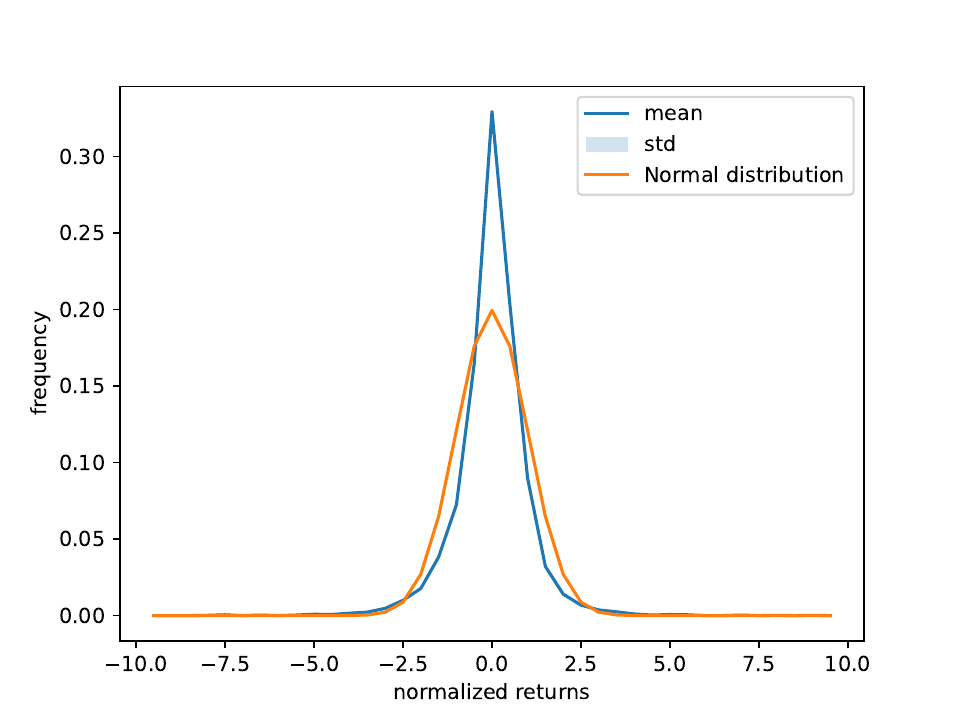}
    \includegraphics[width=0.32\linewidth]{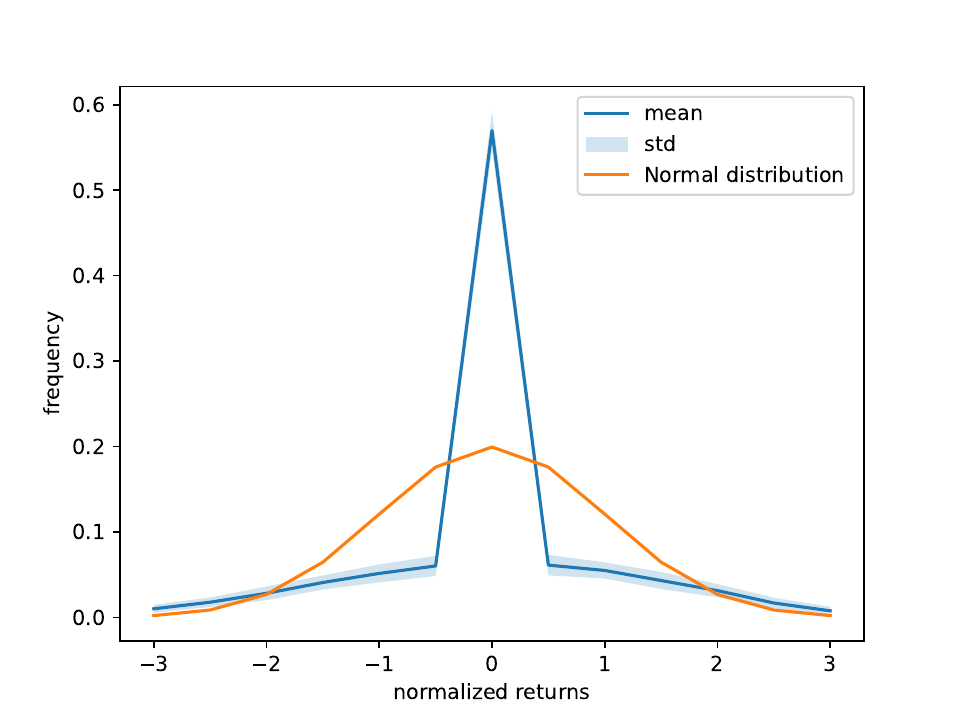}
    \includegraphics[width=0.32\linewidth]{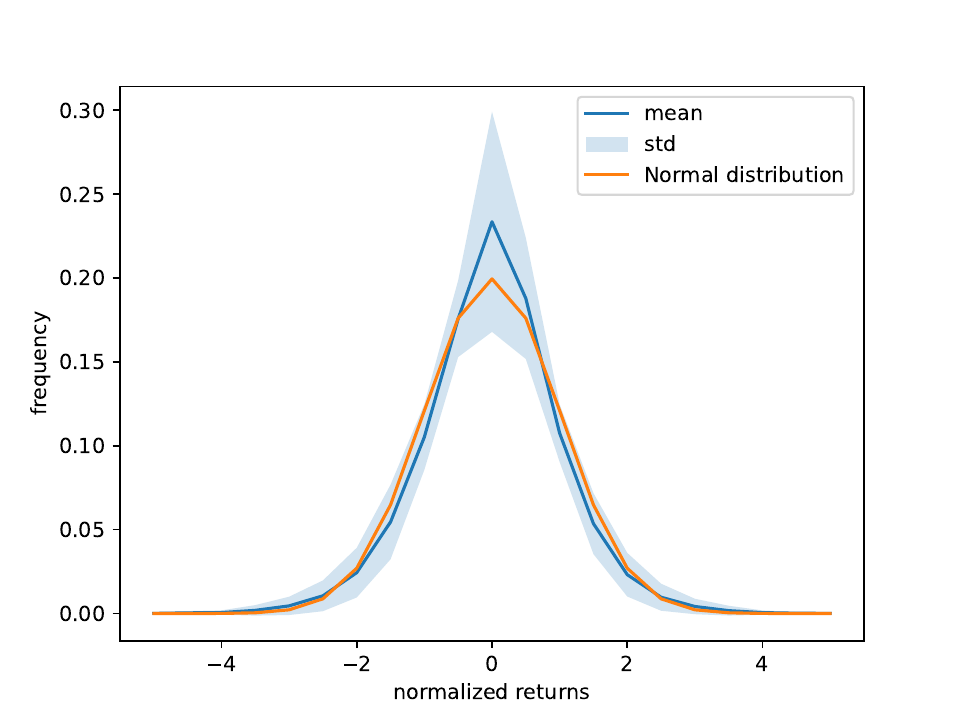}\\
    S\&P500 \hspace{0.17\linewidth} Artificial Market Simulation \hspace{0.17\linewidth} Heston\\
    \includegraphics[width=0.32\linewidth]{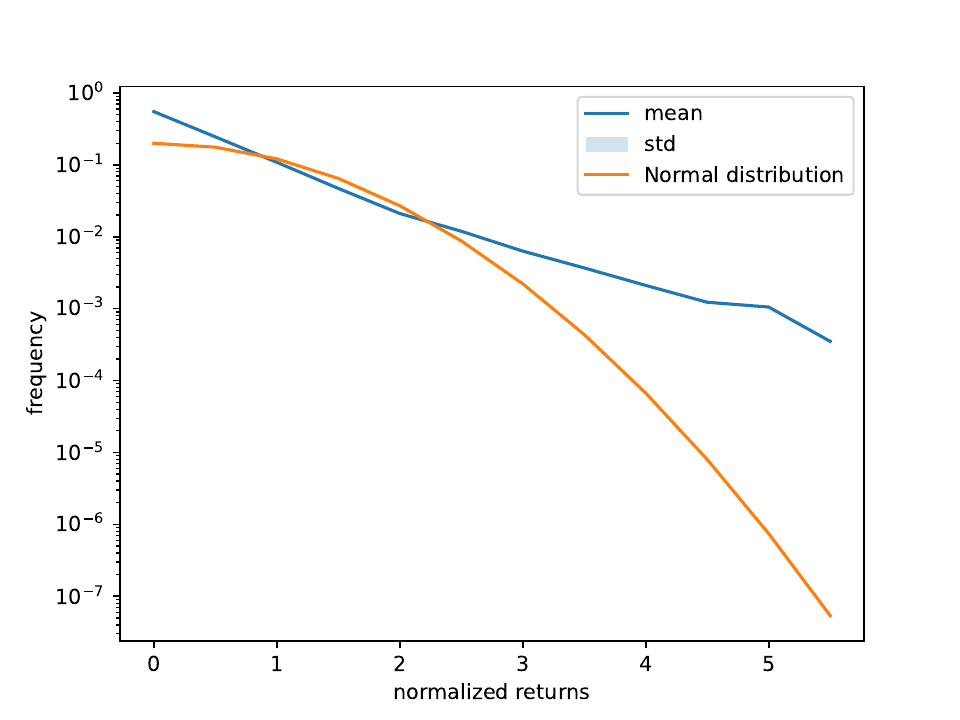}
    \includegraphics[width=0.32\linewidth]{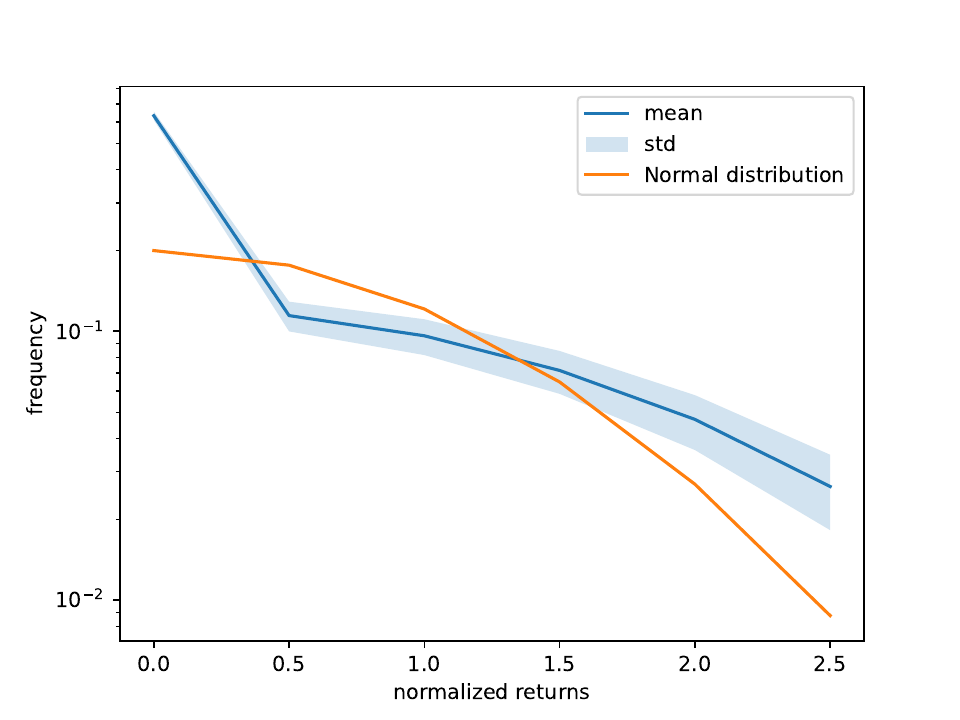}
    \includegraphics[width=0.32\linewidth]{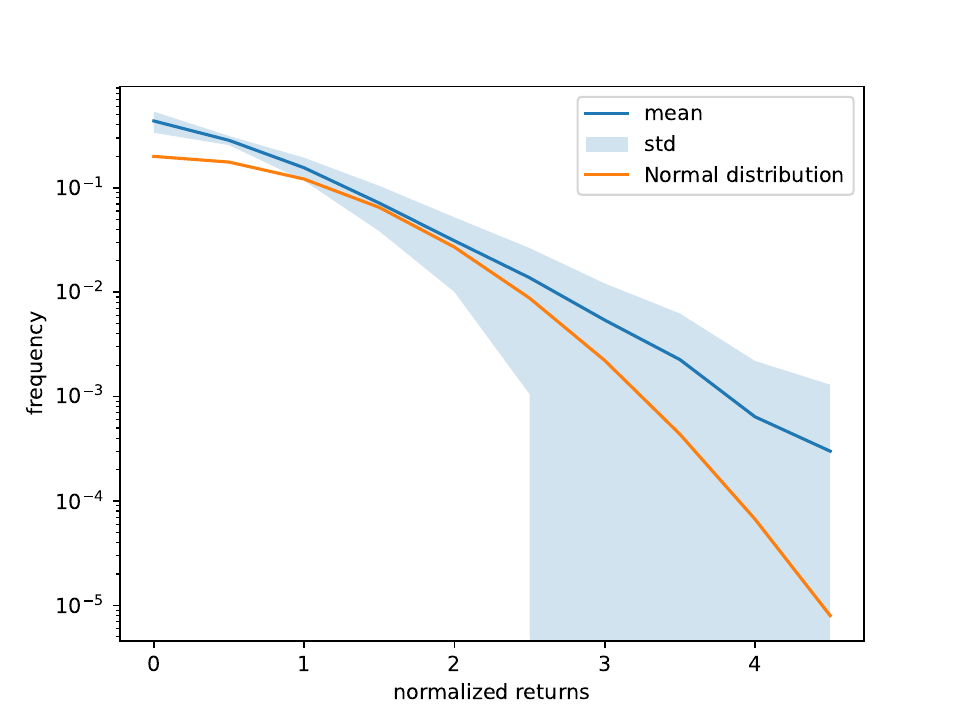}\\
    S\&P500 (absolute) \hspace{0.09\linewidth} Artificial Market Simulation (absolute) \hspace{0.09\linewidth} Heston (absolute)\\
    \caption{Distribution of normalized returns. From left to right, S\&P 500, Artificial Markets, and Heston plotted in 0.5 intervals. Orange indicates a Gaussian distribution. The distribution for the artificial market is based on the case of CVaR ($\alpha=0.95$) for the European Option.}
    \label{fig:returns}
\end{figure*}

Moreover, the results of artificial market simulations are not always the best and are sometimes worse than those of the traditional approach, especially when we employ ERM ($\lambda=10$) or CVaR ($\alpha=0.99$) as the utility function.
This is because artificial market simulations involve tail-risk events that occur more frequently than the traditional approach.
The situations of ERM ($\lambda=10$) and CVaR ($\alpha=0.99$) are very risk-sensitive, and tail risk events have a significant impact on the performance.
Therefore, through the tuning process, artificial market simulations occur more frequently, causing tail-risk events in which the deep hedging model is exposed to greater risk.
However, those very severe risk events occasionally occur in actual markets.
Therefore, the final results depend heavily on whether such severe risk events occur in actual markets.
To investigate more deeply, we plot the distribution of returns in figure \ref{fig:returns}.
According to figure\ref{fig:returns}, the distribution of returns for the artificial market has an exceptionally long tail compared to other models (Note that the x-axis range is far different from the others).
The S\&P 500 also has a tail in its distribution; however, the tails are not as long as those of the artificial market simulations.
However, a high-risk environment may yield better results in learning deep hedging; nevertheless, the details should be investigated in future studies.

\begin{table*}[htb]
    \caption{Parameters of the best-performing models in each setting using the artificial market simulations}
    %\vspace{-2mm}
    \label{tab:tuning-results-sim}
    \centering
    \begin{tabular}{c||c|c|c|c|c||c|c|c|c|c}
        \hline
        Derivative                        & \multicolumn{5}{c||}{European Option} & \multicolumn{5}{c}{Lookback Option}                                                                                                                                                    \\\hline
        \multirow{2}{*}{Utility Function} & \multicolumn{2}{c|}{ERM}              & \multicolumn{3}{c||}{CVaR}          & \multicolumn{2}{c|}{ERM} & \multicolumn{3}{c}{CVaR}                                                                                              \\
                                          & $\lambda=1$                           & $\lambda=10$                        & $\alpha=0.90$            & $\alpha=0.95$            & $\alpha=0.99$ & $\lambda=1$ & $\lambda=10$ & $\alpha=0.90$ & $\alpha=0.95$ & $\alpha=0.99$ \\\hline\hline
        Learning rate                     & $10^{-1}$                             & $10^{-5}$                           & $10^{-2}$                & $10^{-2}$                & $10^{-3}$     & $10^{-3}$   & $10^{-3}$    & $10^{-5}$     & $10^{-1}$     & $10^{-3}$     \\
        $N_{\mathrm{agent/step}}$         & $1$                                   & $10$                                & $1$                      & $10$                     & $10$          & $10$        & $5$          & $5$           & $10$          & $5$           \\
        $w_F$                             & $5.0$                                 & $5.0$                               & $0.0$                    & $50.0$                   & $10.0$        & $0.0$       & $30.0$       & $5.0$         & $1.0$         & $3.0$         \\
        $w_C$                             & $50.0$                                & $50.0$                              & $3.0$                    & $3.0$                    & $1.0$         & $3.0$       & $10.0$       & $1.0$         & $0.0$         & $30.0$        \\
        $\sigma^{*}$                      & $10^{-3}$                             & $10^{-2}$                           & $10^{-2}$                & $10^{-4}$                & $10^{-4}$     & $10^{-4}$   & $10^{-2}$    & $10^{-4}$     & $10^{-4}$     & $10^{-4}$     \\
        $\sigma$                          & $10^{-3}$                             & $10^{-2}$                           & $10^{-3}$                & $10^{-4}$                & $10^{-3}$     & $10^{-3}$   & $10^{-3}$    & $10^{-3}$     & $10^{-2}$     & $10^{-2}$     \\
        $\tau^{*}_{\mathrm{min}}$         & $10^4$                                & $10^5$                              & $10^4$                   & $10$                     & $10$          & $10^5$      & $10^4$       & $10^2$        & $10^4$        & $10^4$        \\
        $\tau^{*}_{\mathrm{max}}$         & $10^4$                                & $10^5$                              & $10^4$                   & $10^5$                   & $10^2$        & $10^5$      & $10^5$       & $10^3$        & $10^5$        & $10^5$        \\
        $\tau_{\mathrm{min}}$             & $10^2$                                & $10^2$                              & $1$                      & $10^2$                   & $1$           & $1$         & $10$         & $10$          & $10$          & $10^2$        \\
        $\tau_{\mathrm{max}}$             & $10^3$                                & $10^2$                              & $1$                      & $10^2$                   & $10$          & $1$         & $10$         & $10$          & $10$          & $10^2$        \\
        $k_{\mathrm{min}}$                & $0.05$                                & $0.10$                              & $0.0$                    & $0.15$                   & $0.15$        & $0.15$      & $0.20$       & $0.20$        & $0.10$        & $0.10$        \\
        $k_{\mathrm{max}}$                & $0.10$                                & $0.15$                              & $0.20$                   & $0.20$                   & $0.15$        & $0.20$      & $0.20$       & $0.20$        & $0.15$        & $0.20$        \\
    \end{tabular}
\end{table*}
\begin{table*}[htb]
    \caption{Parameters of the best-performing models in each setting using Brownian motion}
    %\vspace{-2mm}
    \label{tab:tuning-results-brownian}
    \centering
    \begin{tabular}{c||c|c|c|c|c||c|c|c|c|c}
        \hline
        Derivative                        & \multicolumn{5}{c||}{European Option} & \multicolumn{5}{c}{Lookback Option}                                                                                                                                                    \\\hline
        \multirow{2}{*}{Utility Function} & \multicolumn{2}{c|}{ERM}              & \multicolumn{3}{c||}{CVaR}          & \multicolumn{2}{c|}{ERM} & \multicolumn{3}{c}{CVaR}                                                                                              \\
                                          & $\lambda=1$                           & $\lambda=10$                        & $\alpha=0.90$            & $\alpha=0.95$            & $\alpha=0.99$ & $\lambda=1$ & $\lambda=10$ & $\alpha=0.90$ & $\alpha=0.95$ & $\alpha=0.99$ \\\hline\hline
        Learning rate                     & $10^{-3}$                             & $10^{-3}$                           & $10^{-4}$                & $10^{-3}$                & $10^{-5}$     & $10^{-5}$   & $10^{-1}$    & $10^{-4}$     & $10^{-2}$     & $10^{-3}$     \\
        $\mu$                             & $0.00$                                & $-0.05$                             & $-0.05$                  & $-0.10$                  & $-0.05$       & $0.0$       & $-0.05$      & $0.0$         & $-0.05$       & $0.05$        \\
        $\sigma$                          & $0.15$                                & $0.20$                              & $0.25$                   & $0.50$                   & $0.50$        & $0.15$      & $0.05$       & $0.35$        & $0.05$        & $0.35$        \\
    \end{tabular}
\end{table*}
\begin{table*}[htb]
    \caption{Parameters of the best-performing models in each setting using the Heston model}
    \label{tab:tuning-results-heston}
    %\vspace{-2mm}
    \centering
    \begin{tabular}{c||c|c|c|c|c||c|c|c|c|c}
        \hline
        Derivative                        & \multicolumn{5}{c||}{European Option} & \multicolumn{5}{c}{Lookback Option}                                                                                                                                                    \\\hline
        \multirow{2}{*}{Utility Function} & \multicolumn{2}{c|}{ERM}              & \multicolumn{3}{c||}{CVaR}          & \multicolumn{2}{c|}{ERM} & \multicolumn{3}{c}{CVaR}                                                                                              \\
                                          & $\lambda=1$                           & $\lambda=10$                        & $\alpha=0.90$            & $\alpha=0.95$            & $\alpha=0.99$ & $\lambda=1$ & $\lambda=10$ & $\alpha=0.90$ & $\alpha=0.95$ & $\alpha=0.99$ \\\hline\hline
        Learning rate                     & $10^{-4}$                             & $10^{-3}$                           & $10^{-5}$                & $10^{-4}$                & $10^{-5}$     & $10^{-2}$   & $10^{-5}$    & $10^{-1}$     & $10^{-2}$     & $10^{-3}$     \\
        $\sigma$                          & $0.35$                                & $0.45$                              & $0.30$                   & $0.40$                   & $0.30$        & $0.05$      & $0.25$       & $0.15$        & $0.10$        & $0.15$        \\
        $\kappa$                          & $0.30$                                & $0.40$                              & $0.45$                   & $0.05$                   & $0.05$        & $0.40$      & $0.20$       & $0.45$        & $0.15$        & $0.20$        \\
        $\rho$                            & $1.00$                                & $0.05$                              & $-0.95$                  & $-1.00$                  & $0.30$        & $0.30$      & $-0.35$      & $-0.35$       & $-0.95$       & $0.30$        \\
    \end{tabular}
\end{table*}

In Tables \ref{tab:tuning-results-sim}, \ref{tab:tuning-results-brownian}, and \ref{tab:tuning-results-heston}, we show the parameters of the best-performing models in each setting using artificial market simulations, Brownian motion, and Heston model, respectively.
According to the results, the parameters of the best-performing models differed for each setting.
There are two possible reasons for this.
First, the parameters were not fully optimized during the tuning process.
However, even in the case of Brownian motion, the parameters are not always the same in each setting, even though the number of parameters is extremely small.
Therefore, this is insufficient to explain this difference.
Second, the optimal parameters depend strongly on the situation.
The utility function is particularly vital for deep hedging and risk awareness.
Therefore, for risk-sensitive utility functions such as CVaR ($\alpha=0.99$), high-risk situations such as volatile markets are preferred for deep hedging training.
On the contrary, for risk-insensitive utility functions such as ERM ($\lambda=1$), stable markets are preferred.
Therefore, the best parameters depend highly on the situation and the utility function.
This finding is supported by a previous study \cite{Hirano2023-icaif}.
Thus, the best situation for deep hedging is highly dependent on the utility function, and the tuning process is important for achieving the best performance under each setting.

\begin{figure*}[htb]
    \centering
    \includegraphics[width=0.32\linewidth]{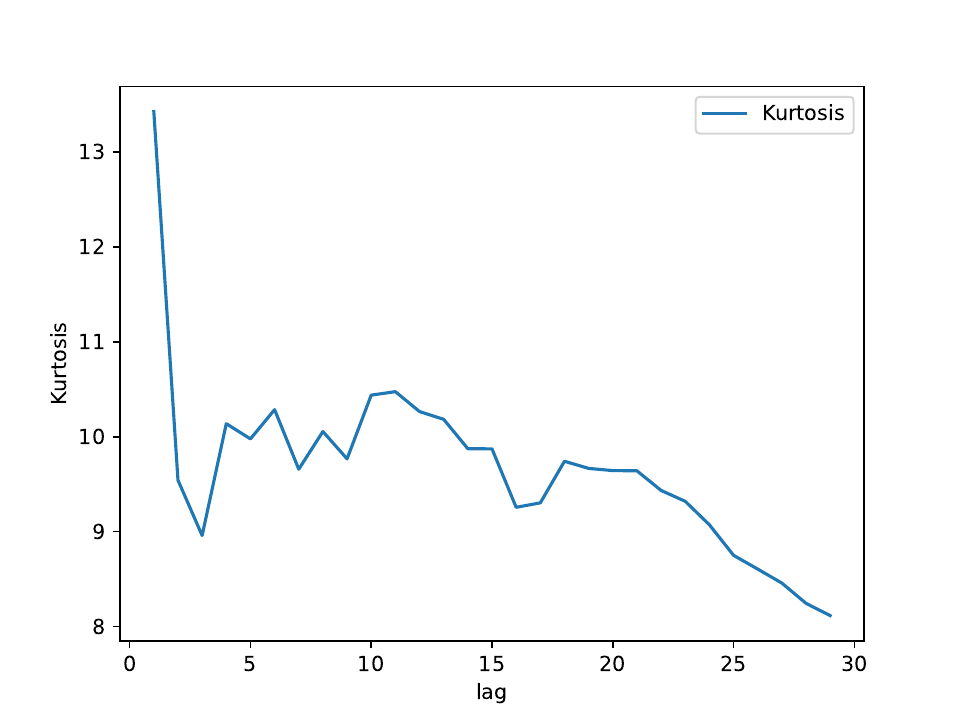}
    \includegraphics[width=0.32\linewidth]{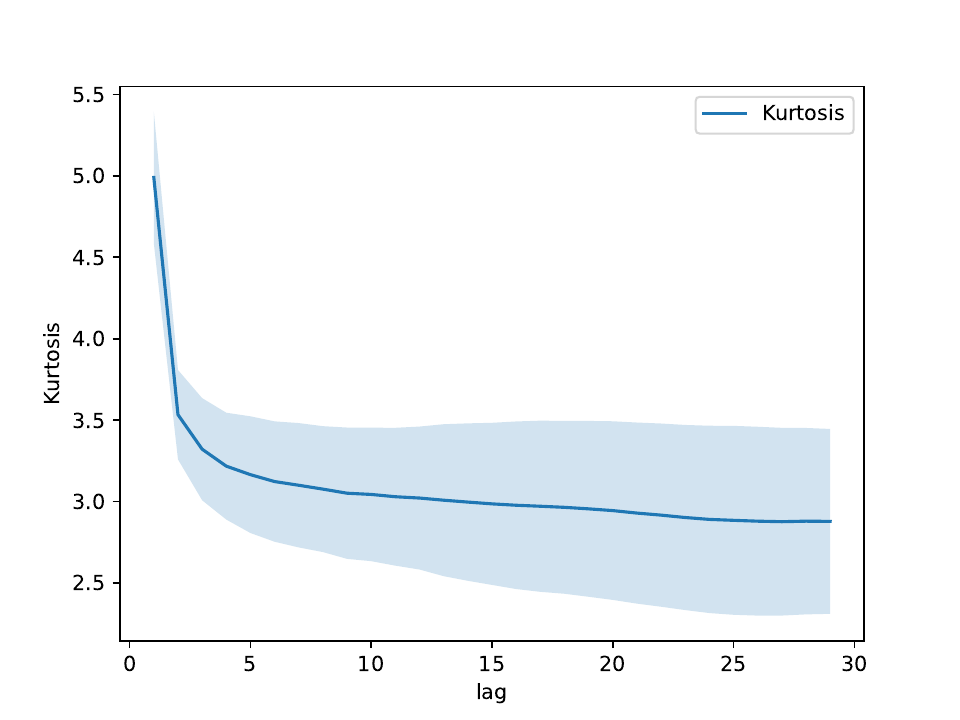}
    \includegraphics[width=0.32\linewidth]{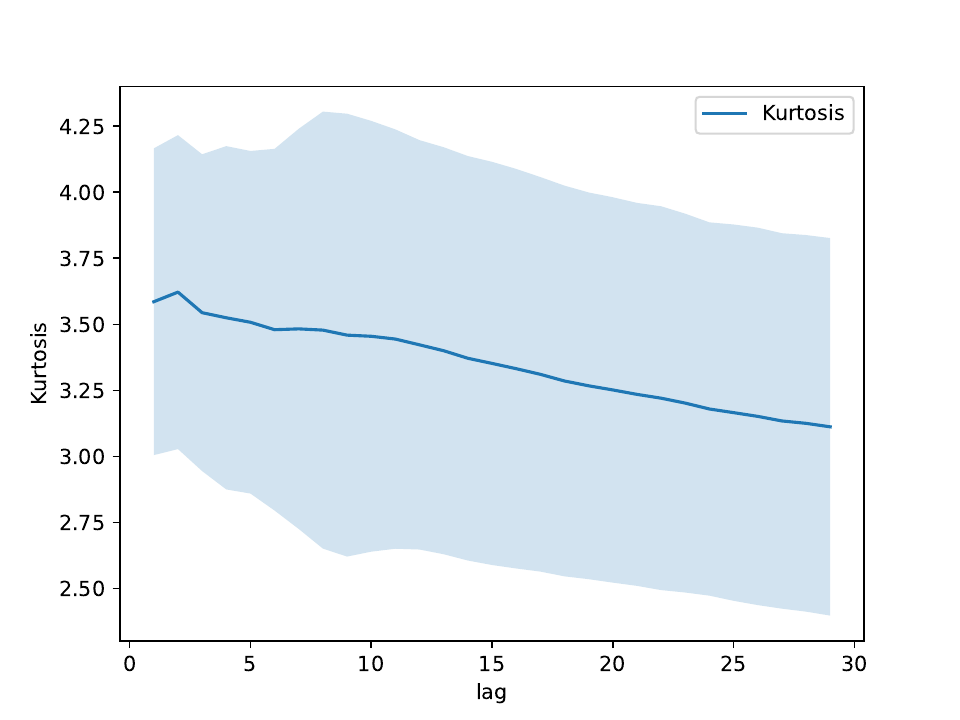}\\
    S\&P500 \hspace{0.17\linewidth} Artificial Market Simulation \hspace{0.17\linewidth} Heston
    \caption{Kutosis of returns. From left to right, S\&P 500, Artificial Markets, and Heston plotted in 0.5 intervals. The horizontal axis is the lag used for the return calculation.}
    \label{fig:kurt}
\end{figure*}

In our proposed method, the artificial market simulation we employ is very simple but outperforms in some settings.
One possible reason for this is that the artificial market simulation can replicate some important features of the actual market that cannot be replicated by other mathematical finance models.
For example, the tail properties of the returns shown in figure \ref{fig:returns} are an example of stylized facts; however, this is also replicated in the Heston model.
However, in terms of kurtosis, that is, the other indices of tail properties, the artificial market simulation replicates better than the Heston model, as shown in figure \ref{fig:kurt}.
In the figure \ref{fig:kurt}, the kurtosis of the Heston model is almost 3, which means that the tails of the returns are almost the same as those in the normal distribution.
However, the artificial market simulation is far higher than 3.0 when the lag used for the return calculation is small.
Even though the maximum kurtosis of the S\&P 500 in the figure is approximately 13, it seemingly better replicates the kurtosis of the actual market data than the Heston model.
Moreover, another stylized fact can be observed in the figure \ref{fig:kurt}.
According to \cite{Cont2001}, aggregational gaussianity is one of the stylized facts of the actual market data.
Aggregational Gaussianity implies that the distribution of the returns becomes more Gaussian as the time scale increases.
According to figure \ref{fig:kurt}, the rate at which the kurtosis of the artificial market simulation decreases is similar to that of the actual market data.
According to these results, artificial market simulation can replicate the stylized facts of actual market data better than the Heston model, and artificial market simulation is a good candidate for comprehending the underlying asset simulation in deep hedging.

This study had some limitations.
First, the artificial market simulation employed in this study is simple and insufficient to replicate actual market data.
Therefore, in the future, we need to develop more sophisticated artificial market simulations for deep hedging and imitate actual market data.
Second, the reason why artificial market simulations outperform in some settings is not fully understood.
Additional investigations are needed to reveal the underlying asset simulation for deep hedging.
Third, the situation that the bias between the training and test data is smaller than in the current setting should be tested.
In this study, the training data are S\&P 500 (old), and the test data are S\&P 500 and BVSP. There is a 50-year difference between the training and test data.
Thus, if we employ rolling training periods, which are very close to the test data period, the performance could be improved.

\section{Conclusion}
This study proposes a new approach using artificial market simulations for underlying asset simulations in deep hedging.
Unlike previous studies, we utilize fully artificial market simulations for underlying asset simulations.
To investigate the effectiveness of the proposed approach, we conducted experiments on option pricing using an actual market simulator and a traditional approach using mathematical finance models.
In the experiments, we tuned certain hyperparameters and compared the results of the proposed approach with those of the traditional approach.
The results show that the proposed approach can achieve almost the same performance as the traditional approach, and, under some settings, the proposed approach can achieve the best performance.
Moreover, we found that the best parameters of the proposed approach were highly dependent on the situation and utility function.
These results show that the artificial market simulation is a good candidate for the underlying asset simulation in deep hedging; however, it still has some limitations.
Future studies should develop more sophisticated artificial market simulations and investigate the robust underlying asset simulations for deep hedging.

\bibliographystyle{IEEEtran}
\bibliography{cite}

\end{document}